\documentclass[11pt,usletter]{article}
\usepackage{jcappub}

\usepackage{graphicx}
\usepackage{hyperref}
\usepackage{color}

\begin{document}

\title{Limits on Entanglement Effects in the String Landscape from Planck and BICEP/Keck Data}

\author{William H.\ Kinney} 
\emailAdd{whkinney@buffalo.edu}
\affiliation{Dept. of Physics, University at Buffalo,
        the State University of New York, Buffalo, NY 14260-1500}

\date{\today}

\abstract{
We consider observational limits on a proposed model of the string landscape in inflation. In this scenario, effects from the decoherence of entangled quantum states in long-wavelength modes in the universe result in modifications to the Friedmann Equation and a corresponding modification to inflationary dynamics. Previous work \cite{Holman:2006an,Holman:2006ny} suggested that such effects could provide an explanation for well-known anomalies in the Cosmic Microwave Background (CMB), such as the lack of power on large scales and the ``cold spot'' seen by both the WMAP and Planck satellites. In this paper, we compute limits on these entanglement effects from the Planck CMB data combined with the BICEP/Keck polarization measurement, and find no evidence for observable modulations to the power spectrum from landscape entanglement, and no sourcing of observable CMB anomalies. The originally proposed model with an exponential potential is ruled out to high significance. Assuming a Starobinsky-type $R^2$ inflation model, which is consistent with CMB constraints, data place a $2\sigma$ lower bound of $b > 6.46 \times 10^7\ {\rm GeV}$ on the Supersymmetry breaking scale associated with entanglement corrections. 
}

\maketitle

\section{Introduction}
\label{sec:Intro}

The so-called ``landscape'' of string theory \cite{Bousso:2000xa,Banks:2003es,Douglas:2003um,Freivogel:2004rd,Denef:2004dm,Dine:2004is,Douglas:2004qg,Douglas:2004zg} remains an intriguing possibility. In the landscape picture, flux compactification results in a discrete collection of a large number of metastable vacua \cite{Bousso:2000xa}, each of which contains independent values of fundamental constants such as the vacuum energy, particle couplings, and Supersymmetry (SUSY) breaking scale. Inflation within these distinct vacua results in a ``multiverse'' of independent universes, each separated from the other by distances large compared to the Hubble length in any one region. Models for inflation in the landscape are numerous \cite{Baumann:2009ni,Burgess:2013sla,Baumann:2014nda}. Despite its appeal from the perspective of fundamental theory, meaningful observational test of the string landscape has remained elusive. Efforts to connect the landscape to observables range from Anthropic arguments \cite{Barrow:1988yia,Susskind:2003kw,Weinstein:2005ef,Ellis:2009gx,Watson:2011vj} to the possibility of bubble collisions between different landscape vacua \cite{Garriga:2006hw,Salem:2012gm,Mersini-Houghton:2014jna,Osborne:2013hea,Wainwright:2013lea,Wainwright:2014pta,Zhang:2015uta,Johnson:2015gma,Johnson:2015mma}, and potential trans-Planckian effects in inflation \cite{Easther:2004vq,Meerburg:2011gd,Ade:2015lrj,Brandenberger:2012aj}. Thus far, no solid evidence for the existence of the string landscape has been discovered. 

One possible mechanism for finding observational effects of the landscape was proposed by Holman, Mersini-Houghton, and Takahashi \cite{Holman:2006an,Holman:2006ny}. Using a mini-superspace approach, they constructed a model of the string landscape as a disordered lattice exhibiting Anderson localization \cite{Kobakhidze:2004gm,MersiniHoughton:2005im}. Localized wavepackets in different landscape vacua exhibit interference on super-Hubble length scales, resulting in nonlocal entanglement between different string vacua. Tracing over superhorizon modes \cite{Halliwell:1984eu,Kiefer:1987ft,Kiefer:1992cn} results in a locally modified Friedmann Equation
\begin{equation}
\label{eq:modifiedFRW}
H^2 = \frac{1}{3 M_{\rm P}^2} \left( \rho + \Delta E_\phi \right),
\end{equation}
where the correction $\Delta E_\phi$ contains the effects of the traced-over superhorizon modes. Here $M_{\rm P} \equiv m_{\rm Pl} / \sqrt{8 \pi} = 2.4357 \times 10^{18}\ {\rm GeV}$ is the reduced Planck mass, and we take a flat Friedmann-Robertson-walker metric
\begin{equation}
ds^2 = dt^2 - a^2(t) d{\bf x}^2
\end{equation}
and $H = \dot a / a$ is the Hubble parameter. This backreaction of long-wavelength modes on the local horizon can result in observable effects. 

In this paper we compare the claimed predictions of this landscape entanglement model with recent Cosmic Microwave Background (CMB) data, in particular the Planck 2015 temperature and polarization spectra \cite{Adam:2015rua,Ade:2015xua,Aghanim:2015xee}, and the Bicep/Keck 2014 combined data \cite{Ade:2015fwj}. The paper is organized as follows: Section \ref{sec:Review} discusses the specific class of landscape models tested. Section \ref{sec:Results} presents methodology and results. Section \ref{sec:Conclusions} presents a summary and conclusions.

\section{Landscape Entanglement Effects in Single-Field Inflation}
\label{sec:Review}

Our approach is phenomenological. To situate landscape entanglement effects in the context of a definite model, we consider single-field inflation, which is an excellent fit to current data \cite{Ade:2015lrj,Ade:2015tva,Martin:2015dha}. We assume a scalar field $\phi$ with a canonical Lagrangian,
\begin{equation}
{\mathcal L} = \frac{1}{2} g^{\mu\nu} \partial_\mu \phi \partial_\nu \phi - V\left(\phi\right),
\end{equation}
where the potential $V\left(\phi\right)$ is a model-dependent function: we consider two examples in this paper, first an exponential potential (Sec. \ref{sec:Exponential}),
\begin{equation}
V\left(\phi\right) = V_0 e^{\lambda \phi / M_{\rm P}},
\end{equation}
and a potential motivated by Starobinsky $R^2$ inflation (Sec. \ref{sec:Starobinsky}),
\begin{equation}
V\left(\phi\right) = V_0 \left(1 - e^{\sqrt{2/3} \phi / M_{\rm P}}\right)^2.
\end{equation}

The energy density and pressure of a homogeneous scalar field are of the perfect fluid form, with
\begin{eqnarray}
\rho &=& \frac{1}{2} \dot\phi^2 + V\left(\phi\right),\cr
p &=& \frac{1}{2} \dot\phi^2 - V\left(\phi\right).
\end{eqnarray}
For a ``slowly rolling'' scalar field, $(1/2)\dot\phi^2 \ll V\left(\phi\right)$, the field exhibits a vacuum-like equation of state, $p \simeq - \rho$, supporting accelerated expansion. The Friedmann equation for the field in the absence of entanglement modifications is
\begin{equation}
H^2 = \frac{1}{3 M_{\rm P}^2} V\left(\phi\right). 
\end{equation}
Entanglement corrections will then introduce a field-dependent modification to the potential,
\begin{equation}
H^2 = \frac{1}{3 M_{\rm P}^2} \left[ V\left(\phi\right) + \delta V\left(\phi\right)\right] \equiv \frac{1}{3 M_{\rm P}^2} V_{\rm eff}\left(\phi\right). 
\end{equation}
Following \cite{Holman:2006an,Holman:2006ny}, introducing entanglement corrections results in an effective potential of the form
\begin{equation}
\label{eq:Veff}
V_{\rm eff}\left(\phi\right) = V\left(\phi\right) + \frac{1}{2}\left(\frac{V\left(\phi\right)}{3 M_{\rm P}^2}\right)^2 \left\vert F\left(\phi\right)\right\vert,
\end{equation}
where
\begin{eqnarray}
\label{eq:F}
F\left(\phi\right) = &&\frac{3}{2} \left(2 + \frac{m^2 M_{\rm P}^2}{V\left(\phi\right)}\right) \ln\left(\frac{3 b^2 M_{\rm P}^2}{V\left(\phi\right)}\right) \cr
&&- \frac{1}{2}\left(1 + \frac{m^2}{b^2}\right) \exp\left(-3 \frac{b^2 M_{\rm P}^2}{V\left(\phi\right)}\right).
\end{eqnarray}
Here $m^2 = V''\left(\phi\right)$ is the scalar field mass, and $b$ is a SUSY-breaking scale associated with the landscape effects, with these effects being suppressed in the limit of high scale, $b \gg 10^{10}\ {\rm GeV}$.  Once the underlying inflationary potential is specified, the entanglement corrections can be calculated as a function of the additional parameter $b$ and can be compared with data. The entanglement correction (\ref{eq:F}) is in fact logarithmically divergent in the limit $b \rightarrow \infty$, which is an artifact of the infrared cutoff on the entanglement scale (see Ref. \cite{Holman:2006an}). The correction should therefore be treated as an effective term, valid only for a limited range of scale. In practice, the logarithmic divergence does not effect constraints on the theory from data. 

Previous work suggested a number of measurable effects from the presence of landscape entanglement corrections to the inflationary potential:
\begin{itemize}
\item{Suppression of the power spectrum at large angular scales, consistent with the ``Axis of Evil'' \cite{deOliveira-Costa:2003utu,Land:2005ad} observed in WMAP data at multipoles $\ell \leq 10$ \cite{Holman:2006ny}. }
\item{Running of the scalar spectral index $n_{\rm S}$ \cite{Holman:2006ny}.}
\item{A suppression of fluctuations due to a discontinuity in the effective potential at a scale determined by the characteristic interference length of the landscape corrections, estimated to be at a wavenumber of $k = 20 h\ {\rm Mpc^{-1} }$, corresponding to a scale of around $200\ {\rm Mpc}$ at a redshift $z \sim 1$ \cite{Holman:2006ny}. This would result in cosmic voids consistent with the observed ``cold spot'' in the WMAP data \cite{Bennett:2010jb,Kovacs:2014ooa}.}
\item{Anomalous cosmic ``bulk flow'' \cite{MersiniHoughton:2008rq}, which was suggested by analyses of the kinetic Sunyaev-Zeldovich effect in the WMAP data \cite{Kashlinsky:2008ut,Watkins:2008hf,Kashlinsky:2009dw,Feldman:2009es}.}
\item{A suppression of the matter power spectrum, resulting in a value about 20-30\% below the value for a $\Lambda$CDM cosmological model.}
\end{itemize}
In this paper, we apply recent CMB data from the Planck satellite and the BICEP/Keck telescopes to obtain limits on power-spectrum effects from landscape entanglement, in particular:
\begin{itemize}
\item{Suppression of power at large angular scale.}
\item{Running of the scalar spectral index.}
\item{Features in the power spectrum consistent with the existence of anomalous structures such as the WMAP/Planck cold spot.}
\item{Suppressed $\sigma_8$ relative to $\Lambda$CDM cosmology.}
\end{itemize}
The latest data strongly constrain scale-dependent modulations of the power spectrum, and we find that the primordial power spectra, including entanglement corrections during inflation, are constrained to be extremely close to a corresponding best-fit $\Lambda$CDM cosmology.  We find no evidence for observably large entanglement corrections in the Planck/BICEP/Keck primordial power spectra, and we obtain a strong lower-bound on the SUSY-breaking scale $b$ for all models considered.

\section{Methodology and Results}
\label{sec:Results}

\begin{figure}
\begin{center}
\includegraphics[width=0.8\textwidth]{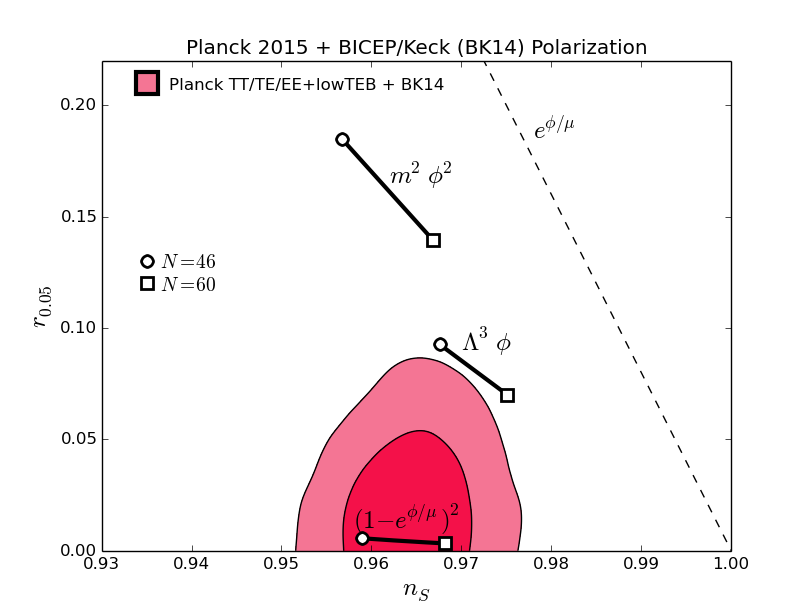}
\end{center}
\caption{Planck + BKP limits on inflationary models in the parameter space of scalar spectral index $n_{\rm S}$ and tensor/scalar ratio $r$, including the models considered in this paper. The pivot scale is taken to be $k_* = 0.05\ {\rm h\ Mpc}^{-1}$.}
\label{fig:PlanckConstraints}
\end{figure}

In this section, we describe the methodology used for constraining landscape entanglement effects on the power spectrum using CMB data. To enable a definite calculation, it is necessary to specify the underlying inflationary potential $V\left(\phi\right)$, since different potentials make different predictions for CMB observables \cite{Dodelson:1997hr,Kinney:1998md}. Single-field inflation predicts approximately power-law spectra $P_\zeta$ for curvature perturbations, and $P_{\rm T}$ for tensor perturbations on observable scales, 
\begin{equation}
P_\zeta\left(k\right) \propto k^{n_{\rm S} - 1},
\end{equation}
and
\begin{equation}
P_{\rm T}\left(k\right) \propto k^{n_T}.
\end{equation}
(See, {\it e.g.}, Ref. \cite{Kinney:2009vz} for a review. See also Refs. \cite{Mukhanov:1981xt,Mukhanov:2003xw} for a discussion of the general predictions of slow-roll inflation.) Both spectra are potential-dependent. It is conventional to describe the parameter space of inflationary observables by the scalar spectral index $n_{\rm S}$ and tensor/scalar ratio $r$, defined as the ratio of the perturbation spectra evaluated at a pivot scale $k = k_*$,
\begin{equation}
r \equiv \frac{P_{\rm T}}{P_\zeta}\bigg\vert_{k = k_*}.
\end{equation}
Different choices of single-field potential can then be compared to constraints from CMB data plotted in the $(n,r)$ plane. Figure \ref{fig:PlanckConstraints} shows a few representative inflation models, compared with the constraints from the Planck 2015 TT/TE/EE+lowTEB temperature and polarization data \cite{Adam:2015rua,Ade:2015xua,Aghanim:2015xee}, and the Bicep/Keck 2014 combined polarization data \cite{Ade:2015fwj}. 
The allowed contours are calculated numerically using a Markov Chain Monte Carlo method with the cosmomc software package \cite{Lewis:2002ah}. For the fiducial best-fit  inflation model, we vary the following background cosmological parameters, for a total fit to eight cosmological parameters\footnote{The data sets themselves contain multiple internal parameters, which we do not list here.}:
\begin{itemize}
\item{Baryon density $\Omega_{\rm b} h^2$.}
\item{Dark matter density $\Omega_{\rm C} h^2$.}
\item{Angular scale of acoustic horizon $\theta$.}
\item{Reionization optical depth $\tau$.}
\item{Hubble parameter $H_0$.}
\item{Power spectrum normalization $A_s$.}
\item{Tensor/scalar ratio $r$.}
\item{Scalar spectral index $n_{\rm S}$.}
\end{itemize}
Curvature $\Omega_{\rm k}$ is set to zero, and the Dark Energy equation of state is fixed at $w = -1$. We fix the number of neutrino species at $N_\nu = 3.046$, with one massive neutrino with mass $m_\nu = 0.06\ {\rm eV}$.  For these constraints, we run 8 parallel chains with Metropolis-Hastings sampling, and use a convergence criterion of the Gelman and Ruben $R$ parameter of $R - 1 < 0.1$. The best-fit model from this set of parameters had a log likelihood of $-\ln({\mathcal L}) = 6794.396$: best-fit parameters are listed in Appendix (\ref{sec:Appendix}). Goodness of fit of models including entanglement corrections are computed relative to this fiducial model, which we label `$\Lambda$CDM+r'. In Sec. (\ref{sec:Exponential}), we consider the exponential model used in Ref. \cite{Holman:2006ny}. In Sec. (\ref{sec:Starobinsky}), we consider entanglement corrections to a Starobinsky $R^2$ model. 

\subsection{Exponential Potential}
\label{sec:Exponential}

We first consider the case of an exponential potential, which was the case considered in Ref. \cite{Holman:2006ny}. We take a base inflationary potential
\begin{equation}
\label{eq:V0exponential}
V\left(\phi\right) = V_0 \exp\left(-\frac{\lambda \phi}{M_{\rm P}}\right). 
\end{equation}
Such a potential is useful because of its analytic simplicity, allowing for exact solutions to the equations for both background dynamics and perturbations. As such, it is a useful test case. However, it has a number of shortcomings from a theoretical perspective: such potentials are not generic to Supergravity (SUGRA) scenarios, although they can be realized in special cases (see for example Ref. \cite{Nakayama:2010ga}). In addition the model lacks a graceful exit from inflation, and the unmodified potential is a poor fit to the Planck data due to overproduction of tensor perturbations. We include the model here for continuity with Refs. \cite{Holman:2006an,Holman:2006ny}. The slow roll parameters are given by
\begin{equation}
\label{eq:epsilon}
\epsilon = \frac{M_{\rm P}}{2} \left(\frac{V'}{V}\right)^2 = \frac{\lambda^2}{2},
\end{equation}
and
\begin{equation}
\label{eq:eta}
\eta = M_{\rm P}^2 \left[\frac{V''}{V} - \frac{1}{2} \left(\frac{V'}{V}\right)^2\right] = \epsilon = \frac{\lambda^2}{2}. 
\end{equation}
The curvature power spectrum is then a pure power-law,
\begin{equation}
P_\zeta\left(k\right) = P_* \left(\frac{k}{k_*}\right)^{n_{\rm S} - 1},
\end{equation}
with
\begin{equation}
n_{\rm S} - 1 = 2 \eta - 4 \epsilon = - \lambda^2.
\end{equation}
Since $\epsilon$ is exactly constant, inflation continues forever in the absence of any new physics at some field value, and we are therefore free to fix $\phi = 0$ at the times when perturbations of wavelength $k = k_* = 0.002\ h^{-1}\ {\rm MpC}$ exited the horizon, with Planck normalization setting $P_* = 2.2 \times 10^{9}$ \cite{Ade:2015xua}. We can then write the power spectrum as a function of field value,
\begin{equation}
\label{eq:Pphi}
P_\zeta\left[\phi\left(k\right)\right] = \frac{1}{24 \pi^2 M_{\rm P}^6} \frac{V\left[\phi\left(k\right)\right]^3}{V'\left[\phi\left(k\right)\right]^2},
\end{equation}
where the prime is a derivative with respect to the field $\phi$, and $\phi\left(k\right)$ is defined as the field value at the time a wavenumber $k$ exited the horizon during inflation. This is straightforward to calculate using
\begin{equation}
d N = - d \ln(k) = - \frac{1}{M_{\rm P} \sqrt{2}} \frac{d \phi}{\sqrt{\epsilon}} = - \frac{d \phi}{\lambda M_{\rm P}}.
\end{equation}
Here 
\begin{equation}
N \equiv - \int{H dt}
\end{equation}
is the number of e-folds of expansion. Then
\begin{equation}
\label{eq:phik0exponential}
\phi_0\left(k\right) = \lambda M_{\rm P} \ln\left(k / k_*\right).
\end{equation}
Inserting this expression into Eq. (\ref{eq:Pphi}) gives an exact power-law,
\begin{equation}
\label{eq:PR}
P_\zeta\left(k\right) = \frac{V_0}{24 \pi^2 \lambda^2 M_{\rm P}^4} \left(\frac{k}{k_*}\right)^{-\lambda^2}.
\end{equation}
Therefore the Planck normalization
\begin{equation}
P_* = \frac{V_0}{24 \pi^2 \lambda^2 M_{\rm P}^4} = 2.2 \times 10^{-9}
\end{equation}
gives a constraint on the scale $V_0$ of inflation,
\begin{equation}
V_0 = \left(5.2 \times 10^{-7}\right) \left(1 - n_{\rm S}\right) M_{\rm P}^4.
\end{equation}
For the Planck best-fit of $1 - n_{\rm S} = 0.04$, this gives
\begin{equation}
V_0 = 2.08 \times 10^{-8}\ M_{\rm P}.
\end{equation}
Exponential potentials also predict substantial tensor production, with
\begin{equation}
\label{eq:PT}
P_{\rm T}\left(k\right) = \frac{V\left[\phi\left(k\right)\right]}{3 \pi^2 M_{\rm P}^4}.
\end{equation}
The ratio $r$ of tensor to scalar power spectra is then
\begin{equation}
r \equiv \frac{P_T\left(k_*\right)}{P_\zeta\left(k_*\right)} = 16 \epsilon = 8 \lambda^2 = 8 \left(1 - n_{\rm S}\right).
\end{equation}
For $1 - n_{\rm S} = 0.04$, this gives $r = 0.32$, which is far higher than the $2\sigma$ upper bound of $r < 0.07$ from Planck TT/TE/EE+lowTEB and BICEP/Keck. Therefore an unmodified exponential potential is inconsistent with CMB data. However, the possibility remains that entanglement corrections may lower the tensor/scalar ratio sufficiently to bring it into agreement with data. 

We consider a modification to the potential of the form (\ref{eq:Veff},\ref{eq:F}), \footnote{There is some ambiguity in the sign of the correction. In Refs. \cite{Holman:2006an,Holman:2006ny}, $F$ is negative and there the correction is negative, but according to the authors, this is a long-overlooked error. (L. Mersini-Houghton and R. Holman, private communication.) Since the exponential potential is an overall poor fit to the data, this does not substantially affect our conclusions. In the case of the Starobinsky potential, the sign of the correction function $F$ is positive, and there is no ambiguity in sign.}
\begin{equation}
V_{\rm eff}\left(\phi\right) = V\left(\phi\right) + f\left[V\left(\phi\right)\right],
\end{equation}
where $V\left(\phi\right)$ is the original potential (\ref{eq:V0exponential}). In general, the modification $f$ depends not only on $V$, but also on $m^2\left(\phi\right) = V''\left(\phi\right)$. Here we approximate the field mass as constant, $m^2 \simeq {\rm const.} = V''\left(\phi_0\right)$, which is a good approximation as long as the field is slowly varying. Then we have
\begin{equation}
V_{\rm eff}'\left(\phi\right) = V'\left(\phi\right) \left[1 + \frac{d f}{d V}\right].
\end{equation}
Then we have
\begin{eqnarray}
d \ln(k) &=& \frac{1}{M_{\rm P} \sqrt{2}} \frac{d \phi}{\sqrt{\epsilon_{\rm eff}}}
= \frac{1}{M_{\rm P} \sqrt{2}} \frac{d \phi}{\sqrt{\epsilon}} \left(\frac{1 + f / V}{1 + df / dV}\right)\cr
&\simeq& \frac{1}{M_{\rm P} \sqrt{2}} \frac{d \phi}{\sqrt{\epsilon}} \frac{V_{\rm eff}}{V},
\end{eqnarray}
which is a good approximation for $f$ slowly varying. The field value as a function of wavenumber for the effective potential is then approximately \cite{Holman:2006ny}\footnote{Refs. \cite{Holman:2006an,Holman:2006ny} evaluate the correction to the field value at fixed $\phi = 0$, but an improved approximation is to evaluate at the uncorrected $\phi\left(k\right)$, given by Eq. (\ref{eq:phik0exponential}).}
\begin{equation}
\label{eq:phikexponential}
\phi(k) \simeq \lambda M_{\rm P} \ln\left(k / k_*\right) \left(\frac{V\left[\phi_0\left(k\right)\right]}{V_{\rm eff}\left[\phi_0\left(k\right)\right]}\right).
\end{equation}
The power spectrum (\ref{eq:Pphi}) can then be calculated as a function of wavenumber $k$ by evaluating the effective potential (\ref{eq:Veff}) and its derivative at $\phi_0\left(k\right)$ given by Eq. (\ref{eq:phikexponential}). We accomplish this numerically via a modified version of the CAMB software (Version: January 2015) \cite{Lewis:1999bs}, with the standard power spectrum code replaced with a direct calculation using the effective potential parameters $V_0$, $\lambda$, and $b$. We constrain these parameters using a Markov Chain Monte Carlo (MCMC) method, as described in Sec. \ref{sec:Results}, with the parameters $n_{\rm S}$ and $r$ replaced by $V_0$, $\lambda$, and $b$, for a total fit to nine cosmological parameters. Sampling in the parameter in $b$ is logarithmic: {\it i.e.}, the parameter directly fit by the MCMC is $\log_{10}(b)$. The tensor power spectrum is also calculated.\footnote{Tensors were neglected in Refs. \cite{Holman:2006an,Holman:2006ny}, a significant omission, since the particular model considered produces a substantial tensor signal.} For each case, we run 16 parallel chains with Metropolis-Hastings sampling, and use a convergence criterion of the Gelman and Ruben $R$ parameter of $R - 1 < 0.15$. 

\begin{figure}
\begin{center}
\includegraphics[width=0.45\textwidth]{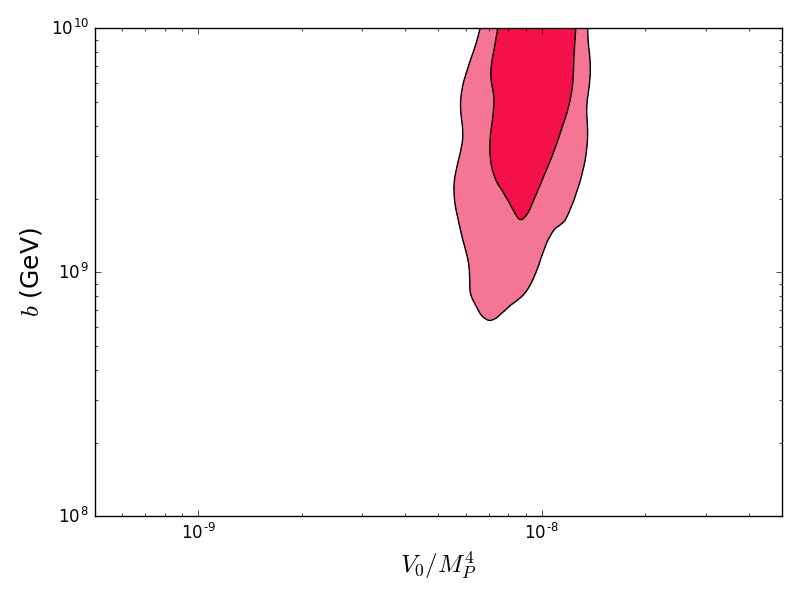}
\includegraphics[width=0.45\textwidth]{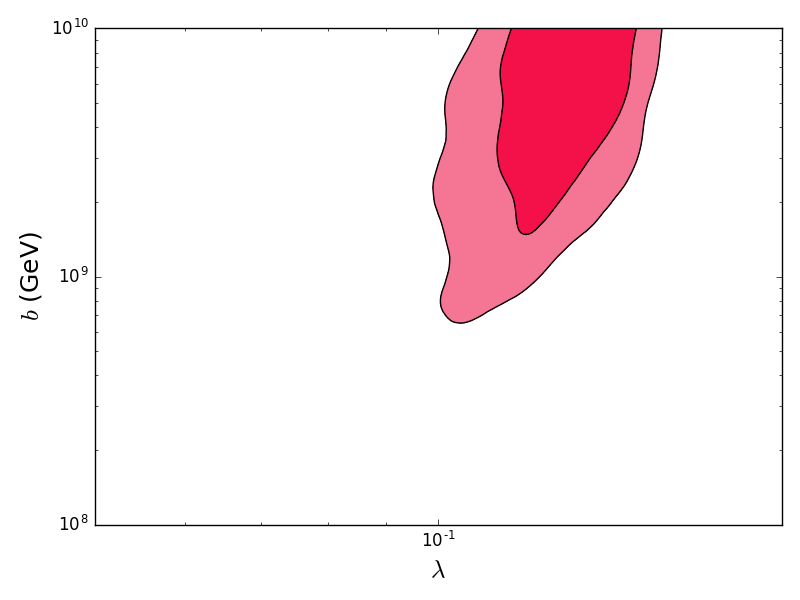}
\includegraphics[width=0.45\textwidth]{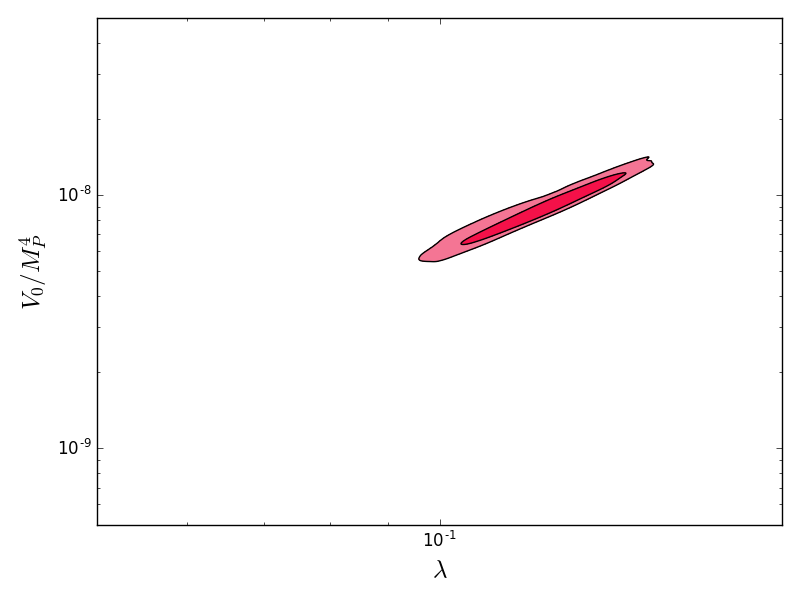}
\end{center}
\caption{Planck limits on the parameters $V_0$, $b$, and $\lambda$ for the Exponential potential. Contours are for 68\% and 95\% confidence limits. These best-fit regions should not be considered allowed regions, as all are a poor statistical fit to Planck data. Plot limits represent actual ranges of parameters scanned by the MCMC.}
\label{fig:Exponential}
\end{figure}

\begin{figure}
\begin{center}
\includegraphics[width=0.45\textwidth]{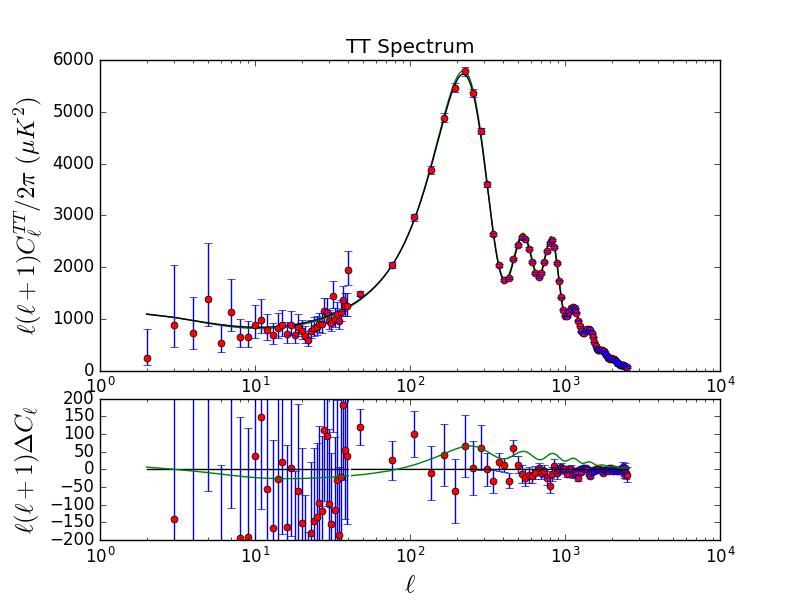}
\includegraphics[width=0.45\textwidth]{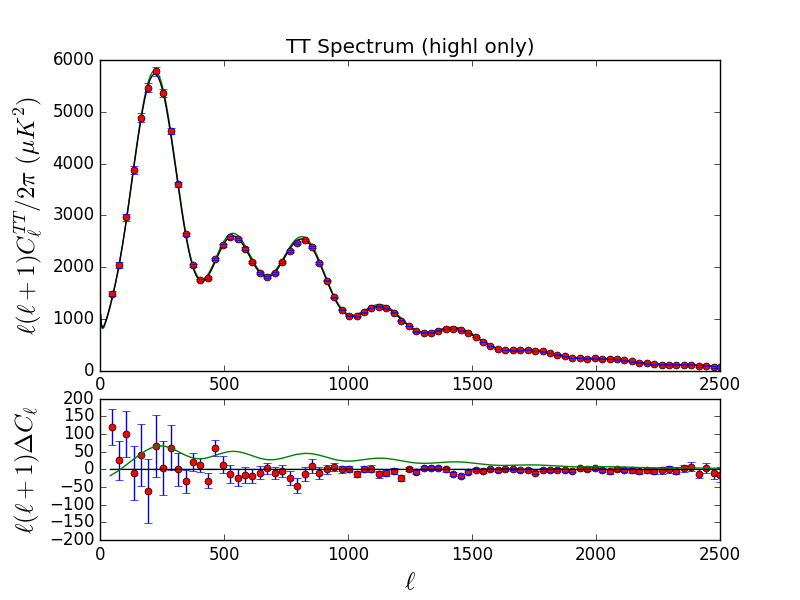}
\includegraphics[width=0.45\textwidth]{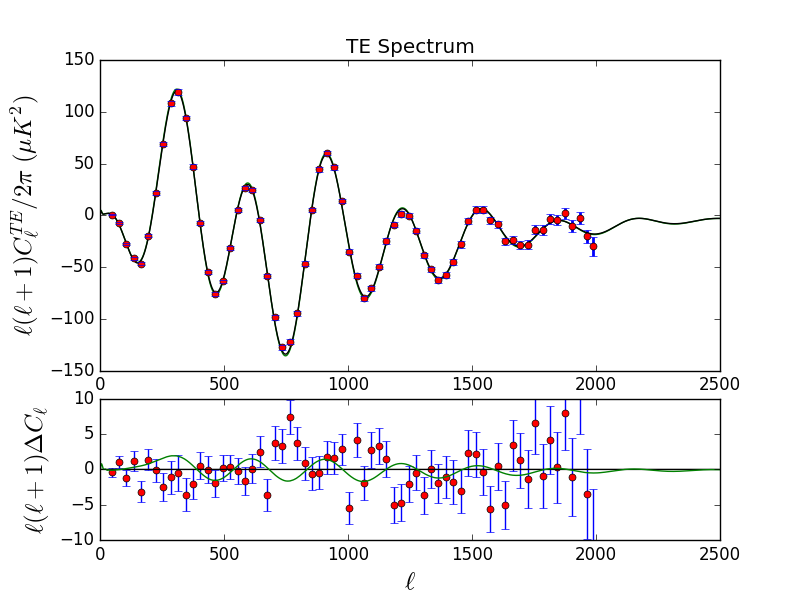}
\includegraphics[width=0.45\textwidth]{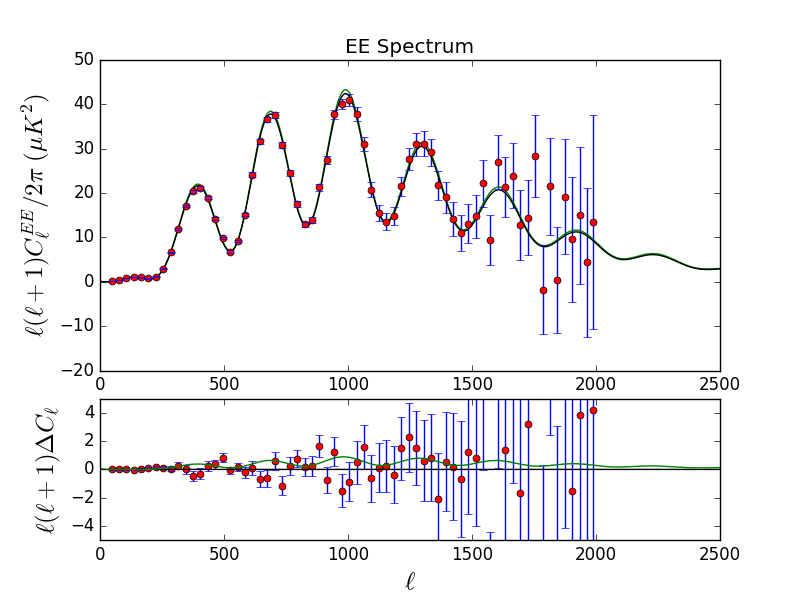}
\end{center}
\caption{CMB angular power spectra for the best-fit Exponential model (green), and the best-fit $\Lambda$CDM+r model (black). Points with error bars are the Planck 2015 data. Bottom pane in each plot shows residuals relative to the $\Lambda$CDM+r best-fit. This can be compared to Fig. 1 of Ref. \cite{Holman:2006ny}}.
\label{fig:ExponentialCl}
\end{figure}

Figure \ref{fig:Exponential} shows constraints on the potential parameters $V_0$, $\lambda$, and $b$. We obtain a lower-bound on the SUSY breaking scale $b$ of $b > 1.32 \times 10^9\ {\rm GeV}$. The overall fit of the model to the data is poor, with a best-fit log likelihood of $- \ln{\mathcal  L} = 6807.6640$, compared with $- \ln{\mathcal L} = 6794.396$ for the $\Lambda$CDM+r model. Figure (\ref{fig:ExponentialCl}) shows the CMB angular power spectra for the $\Lambda$CDM+r model, and the best-fit exponential model.  We conclude that the exponential potential, even with entanglement corrections, is inconsistent with data. 

\subsection{Starobinsky Potential}
\label{sec:Starobinsky}

In this section, we consider a model which is in good agreement with existing data, and place limits on the magnitude of entanglement corrections. For this purpose, we choose the Starobinsky $R^2$ model \cite{Starobinsky:1980te,Kehagias:2013mya}. Despite its fundamental formulation as a modified-gravity theory, Starobinsky inflation can be re-cast in Einstein frame as an effective single-field inflation model, with potential
\begin{equation}
V\left(\phi\right) = V_0 \left[1 - e^{-\phi  / \mu}\right]^2,
\end{equation}
where the mass scale $\mu$ is not a free parameter, but is fixed to be
\begin{equation}
\mu = \sqrt{\frac{2}{3}} M_{\rm P}.
\end{equation}
Starobinsky-type inflation has been embedded into SUGRA scenarios in several recent works \cite{Ketov:2010qz,Ketov:2012jt,Ferrara:2013wka,Ketov:2014qoa,Ferrara:2015ela}, and also appears in Higgs inflation \cite{Bezrukov:2007ep}. As in Section \ref{sec:Exponential}, we wish to calculate an analytic expression for $\phi\left(k\right)$ suitable for numerical calculation of the power spectra. In the case of the Starobinsky potential, this will be approximate instead of exact. 

The slow roll parameters (\ref{eq:epsilon}), (\ref{eq:eta}) are given by:
\begin{equation}
\epsilon = \frac{4}{3} \left(\frac{1}{e^{\phi / \mu} - 1}\right)^2,
\end{equation}
and
\begin{equation}
\eta = \left(\frac{4}{3}\right)  \frac{e^{-\phi / \mu}\left(2 e^{-\phi / \mu} - 1\right) - 1}{\left(e^{-\phi/\mu} - 1\right)^2}.
\end{equation}
The number of e-folds is
\begin{eqnarray}
\label{eq:Nefolds}
d N = - d \ln(k) &=& - \frac{d\phi}{M_{\rm P} \sqrt{2 \epsilon}}\cr &=& \frac{3}{4} \left(e^{\phi / \mu} - 1\right) d\left(\frac{\phi}{\mu}\right).
\end{eqnarray}
For $\phi / \mu \gg 1$, the number of e-folds is then approximately
\begin{equation}
N \simeq \frac{3}{4} e^{\phi/\mu}.
\end{equation}
We then have a lowest-order approximation for $\phi\left(k\right)$, 
\begin{equation}
\label{eq:phi0star}
\phi_0\left(k\right) = \mu \ln{\left[\frac{4}{3}\left(N_* - \ln{\frac{k}{k_*}}\right)\right]},
\end{equation}
where $N_*$ is the number of e-folds at the pivot scale $k = k_*$. For this analysis, we take $N_* = 60$, consistent with high-scale inflation and rapid reheating.\footnote{Due to the extreme flatness of the potential, constraints are not strongly dependent on the details of reheating, which fix $N_*$.} To leading order in $N_*$, the scalar spectral index is
\begin{equation}
n_{\rm S} - 1 \simeq - \frac{2}{N_*} = 0.033,
\end{equation}
and the tensor/scalar ratio is
\begin{equation}
r \simeq \frac{12}{N_*^2} = 0.0033.
\end{equation}
These values are consistent with constraints from Planck (Fig. \ref{fig:PlanckConstraints}). We incorporate the entanglement modifications as with the case of the exponential potential, 
\begin{equation}
\label{eq:phieff0}
e^{\phi\left(k\right) / \mu} \simeq \ln{\left[\frac{4}{3}\left(N_* - \ln{\frac{k}{k_*}}\right)\right]}  \left(\frac{V\left[\phi_0\left(k\right)\right]}{V_{\rm eff}\left[\phi_0\left(k\right)\right]}\right),
\end{equation}
where the effective potential $V_{\rm eff}$ is again derived from $V\left(\phi\right)$ by Eqs. (\ref{eq:Veff}) and (\ref{eq:F}).

\begin{figure}
\begin{center}
\includegraphics[width=0.8\textwidth]{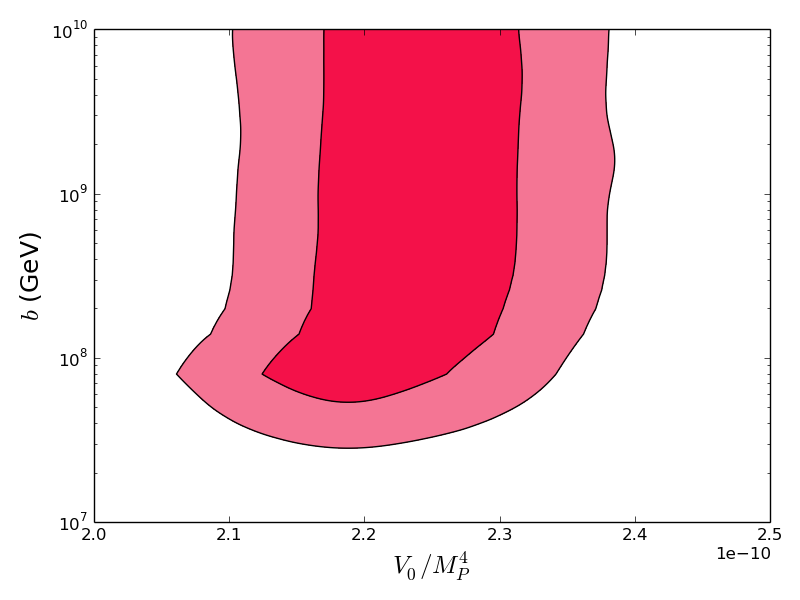}
\end{center}
\caption{Planck 68\% (dark shaded) and 95\% (light shaded) limits on the parameters $V_0$ and $b_{\rm SUSY}$ for the Starobinsky potential. }
\label{fig:Starobinsky}
\end{figure}

\begin{figure}
\begin{center}
\includegraphics[width=0.45\textwidth]{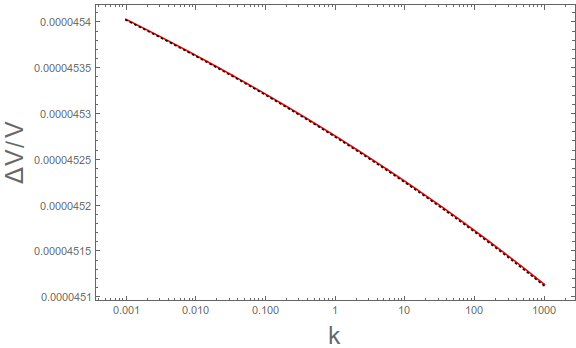}
\includegraphics[width=0.45\textwidth]{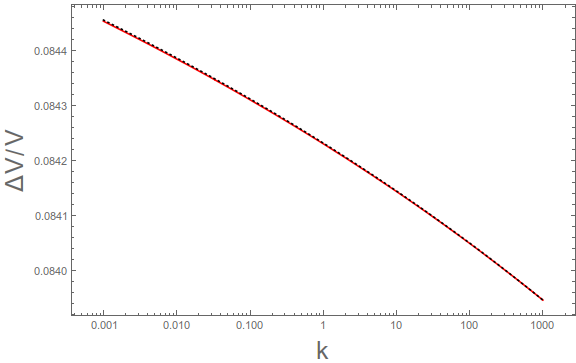}
\end{center}
\caption{$\Delta V / V$ for the best-fit Starobinsky model (left), and the $2\sigma$ extremal model (right). Solid red curve is a numerical evaluation of $\phi\left(k\right)$, dotted black curve is the analytical approximation (\ref{eq:phieff}). $k$ is in units of ${\rm h\ Mpc}^{-1}$.}
\label{fig:StarobinskyDeltaV}
\end{figure}

\begin{figure}
\begin{center}
\includegraphics[width=0.45\textwidth]{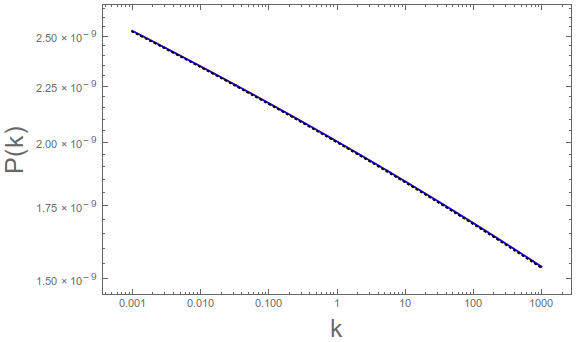}
\includegraphics[width=0.45\textwidth]{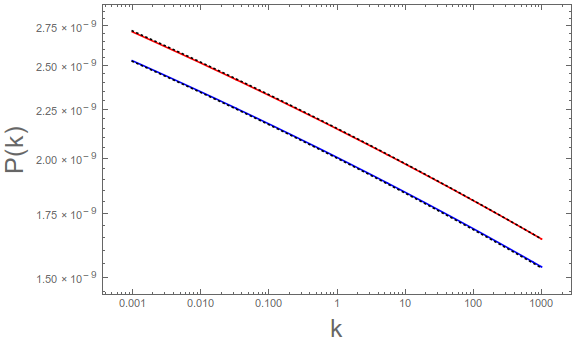}
\end{center}
\caption{$P(k)$ for the best-fit Starobinsky model (left), and the $2\sigma$ extremal model (right). Solid red curve is the power spectrum for the entanglement-modified potential for a numerical evaluation of $\phi\left(k\right)$, and the corresponding dotted black curve is the analytical approximation (\ref{eq:phieff}). The blue curve is the unmodified power spectrum for a numerical evaluation of $\phi\left(k\right)$, and the corresponding dotted black curve is the analytical approximation. Note that in the top plot, the red curve is invisible because it is so close to the unmodified power spectrum.  $k$ is in units of ${\rm h\ Mpc}^{-1}$.}
\label{fig:StarobinskyPk}
\end{figure}

\begin{figure}
\begin{center}
\includegraphics[width=0.45\textwidth]{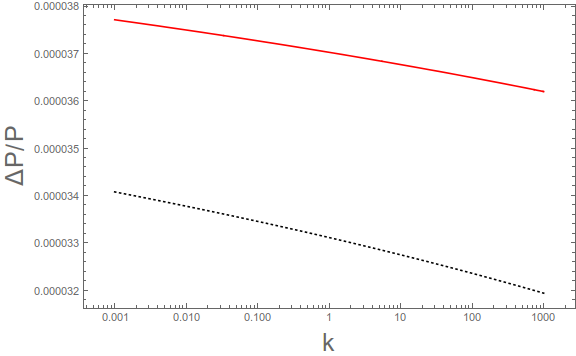}
\includegraphics[width=0.45\textwidth]{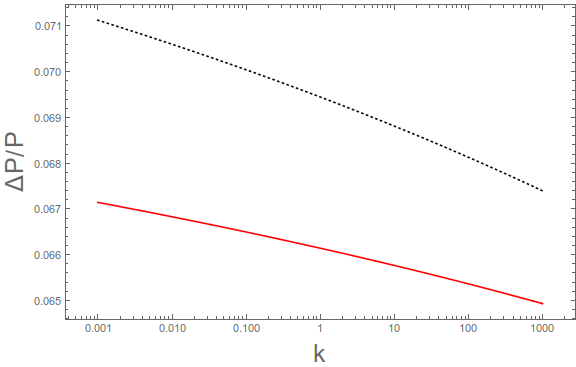}
\end{center}
\caption{$\Delta P(k) / P(k)$ for the best-fit Starobinsky model (left), and the $2\sigma$ extremal model (right). Solid red curve is the power spectrum for the entanglement-modified potential for a numerical evaluation of $\phi\left(k\right)$, and the corresponding dotted black curve is the analytical approximation (\ref{eq:phieff}).  $k$ is in units of ${\rm h\ Mpc}^{-1}$. }
\label{fig:StarobinskyDeltaP}
\end{figure}

\begin{figure}
\begin{center}
\includegraphics[width=0.45\textwidth]{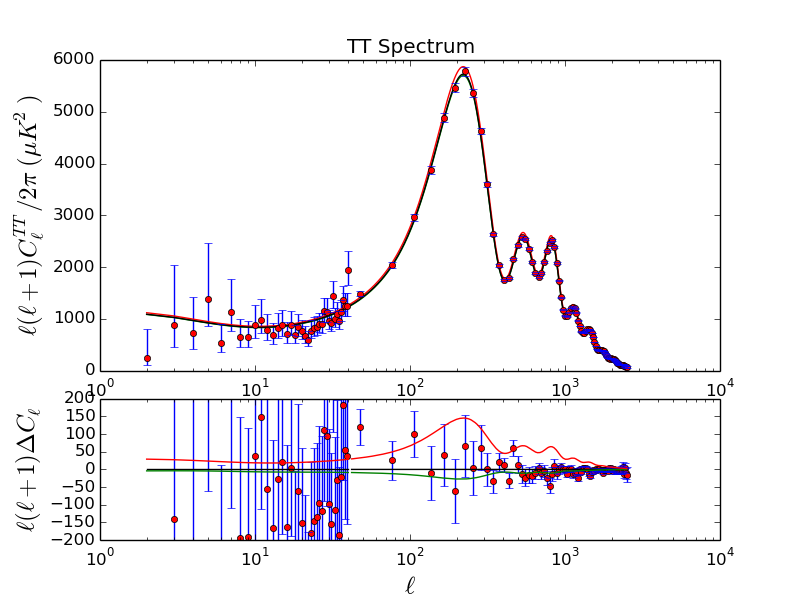}
\includegraphics[width=0.45\textwidth]{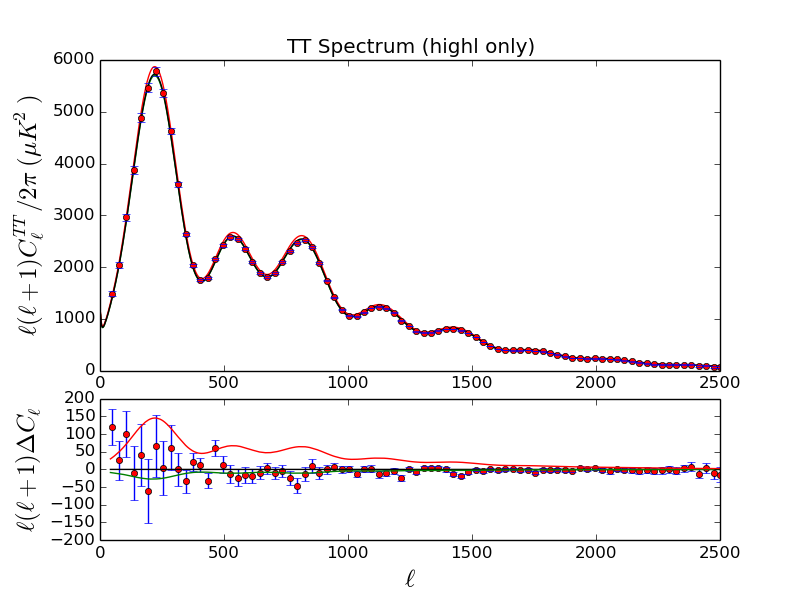}
\includegraphics[width=0.45\textwidth]{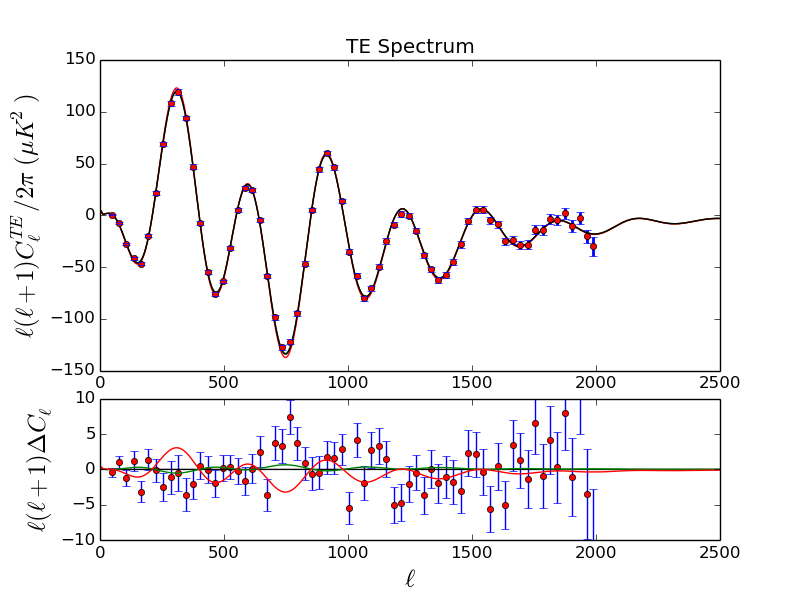}
\includegraphics[width=0.45\textwidth]{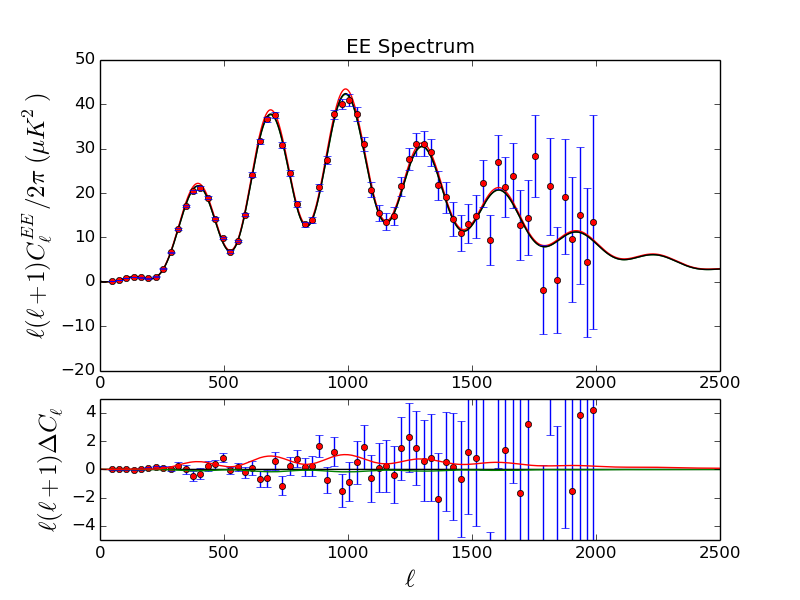}
\end{center}
\caption{CMB angular power spectra for the best-fit Starobinsky model (green), the extremal Starobinsky model (red), and the best-fit $\Lambda$CDM+r model (black). Points with error bars are the Planck 2015 data. Bottom pane in each plot shows residuals relative to the $\Lambda$CDM+r best-fit.}
\label{fig:StarobinskyCl}
\end{figure}

Numerical evaluation of the full slow-roll expression shows, however, that Eqs. (\ref{eq:phi0star}) and (\ref{eq:phieff0}) are accurate only to within about 10\%, insufficient for an accurate numerical evaluation of the power spectrum. We can improve the approximation iteratively as follows: exactly solving the integral (\ref{eq:Nefolds}) gives
\begin{equation}
N = \frac{3}{4}\left(e^x - x\right)\bigg\vert_{\phi_e / \mu}^{\phi\left(N\right) / \mu},
\end{equation}
where $\phi_e$ is the field value at the end of inflation, where $\epsilon\left(\phi_e\right) = 1$, given by
\begin{equation}
\left(\phi_e / \mu\right) = 1 + \frac{2}{\sqrt{3}}.
\end{equation}
The solution for $\phi\left(k\right)$ requires solution of a transcendental equation. We can approximate the solution by inserting the lowest-order solution $\phi_0\left(k\right)$ into the linear term above, with the result
\begin{equation}
\label{eq:phieff}
\phi_{\rm eff}\left(k\right) \simeq \sqrt{\frac{3}{2}} M_{\rm P} \ln{\left[e^{\phi\left(k\right) / \mu}+ \frac{\phi_0\left(k\right)}{\mu} + e^{\phi_e / \mu} - \frac{\phi_e}{\mu}\right]},
\end{equation}
where $\phi\left(k\right)$ is given by Eq. (\ref{eq:phieff0}), and $\phi_0\left(k\right)$ is given by the lowest-order approximation (\ref{eq:phi0star}).\footnote{Here one could replace $\phi_0\left(k\right)$ with $\phi\left(k\right)$ to achieve a potential improvement in the approximation. The difference is in practice negligible, and Eq. (\ref{eq:phieff}) represents the actual implementation in the code.} This approximation is accurate at the level of $\Delta\phi / \phi \sim O(10^{-4})$, relative to a numerical solution.  Primordial power spectra are calculated, as in the case of the exponential potential, using a modified version of CAMB directly implementing Eq. (\ref{eq:PR}) for the scalar power spectrum, and Eq. (\ref{eq:PT}) for the tensor power spectrum. Figure (\ref{fig:Starobinsky}) shows the resulting constraints for the Starobinsky potential in the parameter plane $b$, $V_0$. We obtain a $2\sigma$ lower bound of $b > 6.46 \times 10^7\ {\rm GeV}$. We check consistency by comparing the analytic approximation for $P_\zeta\left(k\right)$ to an exact numerical integration of the slow roll parameters, and find agreement at the $O(0.001)$ level for points within the 95\% confidence allowed region for Planck, demonstrating the consistency of the analytic approximations used in CAMB.  For comparison with data, we consider two models:
\begin{enumerate}
\item{The best-fit Starobinsky model, with $V_0 = 2.229236 \times 10^{-10} M_{\rm P}^4$ and $b = 6.99 \times 10^{8}\ {\rm GeV}$.}
\item{An ``extremal'' Starobinsky model, with $V_0 = 2.229236 \times 10^{-10} M_{\rm P}^4$ and $b = 6.46 \times 10^7\ {\rm GeV}$, just at the 95\%-confidence lower bound for the parameter $b$, leaving all other cosmological parameters fixed at their best-fit values. This is a poor fit to the data (see Fig. \ref{fig:StarobinskyCl}), and is included as a reference to the most extreme modulation which is even marginally consistent with data.}
\end{enumerate}
Figure (\ref{fig:StarobinskyDeltaV}) shows the entanglement modification $\Delta V / V$ to the potential for the best-fit and extremal models, and Figs. (\ref{fig:StarobinskyPk}) and (\ref{fig:StarobinskyDeltaP}) show the corresponding power spectra and the modulation $\Delta P / P$, respectively. In the best-fit case, the power spectrum modulation is negligible, $\Delta P / P \simeq 3 \times 10^{-5}$. In the extremal case, the power spectrum is of order 7\%. Of particular note is the absence of any features or running in the modulated power spectrum, even in the extremal case. Nor is there suppression of power at large scales: in fact, power is slightly \emph{enhanced} at large scales. In particular, we see no evidence of power suppression near the quadrupole, no running of the spectral index, and no features in the power spectrum which would explain observed CMB anomalies such as the WMAP/Planck cold spot. Figure (\ref{fig:StarobinskyCl}) shows the CMB angular power spectra for the $\Lambda$CDM+r model, the best-fit Starobinsky model, and the extremal model. 

\section{Conclusions}
\label{sec:Conclusions}

The prospect of detecting signs of stringy physics or other quantum-gravitational phenomena in high-precision cosmological data remains a tantalizing one, and new precision cosmological data are approaching the required accuracy to test stringy modifications to cosmology. In this paper, we consider one such proposed modification, relying on quantum interference effects between Anderson-localized wavepackets in a ``landscape'' of approximately $10^{500}$ distinct string vacua \cite{Halliwell:1984eu,Kiefer:1987ft,Kiefer:1992cn,Kobakhidze:2004gm,MersiniHoughton:2005im,Holman:2006an,Holman:2006ny}. The authors claim a number of predictions for cosmology:
\begin{itemize}
\item{Suppression of the power spectrum at large angular scales, consistent with the ``axis of evil'' \cite{deOliveira-Costa:2003utu,Land:2005ad} observed in WMAP data at multipoles $\ell \leq 10$ \cite{Holman:2006ny}. }
\item{Running of the scalar spectral index $n_{\rm S}$ \cite{Holman:2006ny}.}
\item{A suppression of fluctuations due to a discontinuity in the effective potential at a scale determined by the characteristic interference length of the landscape corrections, estimated to be at a wavenumber of $k = 20 h\ {\rm Mpc^{-1} }$, corresponding to a scale of around $200\ {\rm Mpc}$ at a redshift $z \sim 1$ \cite{Holman:2006ny}. This would result in cosmic voids consistent with the observed ``cold spot'' in the WMAP and Planck data \cite{Bennett:2010jb}.}
\item{Anomalous cosmic ``bulk flow'' \cite{MersiniHoughton:2008rq}, which was suggested by analyses of the kinetic Sunyaev-Zeldovich effect in the WMAP data \cite{Kashlinsky:2008ut,Watkins:2008hf,Kashlinsky:2009dw,Feldman:2009es}.}
\item{A suppression of the matter power spectrum, resulting in $\sigma_8 \simeq 0.8$, about 20-30\% below a best-fit $\Lambda$CDM figure quoted by the authors of $\sigma_8 = 1.1$ \cite{Holman:2006ny}. (It is unclear where the high value for $\sigma_8$ was obtained, since the WMAP3 best-fit at the time was $\sigma_8 = 0.76 \pm 0.05$ \cite{Spergel:2006hy}.)}
\end{itemize}
A decade after the model was first proposed, new data test almost all of these claims:
\begin{itemize}
\item{The presence of the anomalous CMB ``cold spot'' seen by WMAP was confirmed by Planck  \cite{Ade:2015hxq,Schwarz:2015cma}, possibly associated with an anomalous void at redshift $z = 0.22$ \cite{Kovacs:2014ooa}.}
\item{Power suppression at the quadrupole remains in the Planck data, although the statistical significance of alignments among the lowest multipoles is debated \cite{Schwarz:2015cma,Rassat:2014yna}.}
\item{A matter power spectrum with $\sigma_8 \simeq 0.8$ is favored by both the WMAP data and Planck data. The Planck best-fit $\Lambda$CDM cosmology, without entanglement corrections, has $\sigma_8 = 0.830 \pm 0.015$ \cite{Ade:2015xua}.}
\item{The presence of anomalously large bulk flow velocities has now been convincingly ruled out by a large number of independent measurements \cite{Keisler:2009nw,Osborne:2010mf,Dai:2011xm,Nusser:2011tu,Turnbull:2011ty,Lavaux:2012jb,Ade:2013dsi,Rathaus:2013ut,Ma:2013oja,Huterer:2015gpa,Scrimgeour:2015khj}.
Current evidence strongly disfavors any model predicting large anomalous bulk flow.}
\end{itemize}

In this paper, we use the Planck 2015 temperature/polarization data\cite{Adam:2015rua,Ade:2015xua,Aghanim:2015xee}, and the Bicep/Keck 2014 combined data \cite{Ade:2015fwj} to test the following claimed predictions of the landscape entanglement model:
\begin{itemize}
\item{Suppression of power at large angular scale.}
\item{Running of the scalar spectral index.}
\item{Features in the power spectrum consistent with the existence of anomalous structures such as the WMAP/Planck cold spot.}
\item{Suppressed $\sigma_8$ relative to $\Lambda$CDM cosmology.}
\end{itemize}
We consider two cases: (1) The exponential inflationary potential proposed in the original paper \cite{Holman:2006ny}, and (2) a Starobinsky $R^2$ inflation potential. Both models are compared to a fiducial $\Lambda$CDM + tensor model ($\Lambda$CDM+r) to quantify the relative likelihood of the model. Our conclusions are as follows:
\begin{itemize}
\item{The exponential potential has three parameters: the height of the potential $V_0$, the coupling $\lambda$, and the SUSY-breaking scale $b$. associated with entanglement.  This model is entirely ruled out by the data, with a best-fit likelihood $-\ln{({\mathcal L})} = 6807.6640$, compared with $-\ln{({\mathcal L})} = 6794.3960$ for the $\Lambda$CDM+r model. This is primarily because of overproduction of tensor modes, with a best-fit tensor/scalar ratio of $r = 0.130768$, compared with a $\Lambda$CDM+r 95\%-confidence upper bound of $r < 0.06693180$. There is no evidence for observable entanglement corrections to the primordial power spectrum, with a 95\%-confidence lower-bound of $b > 1.12 \times 10^9\ {\rm GeV}$, if the exponential potential is assumed as a prior.}
\item{The Starobinsky $R^2$ potential has two parameters: the height of the potential $V_0$, and the SUSY-breaking scale $b$. The best-fit Starobinsky $R^2$ model, including entanglement corrections, is a good fit to data, with a best-fit $-\ln{({\mathcal L})} = 6792.9980$. (Remarkably, even the unmodified Starobinsky potential gives a better fit to the data with \emph{fewer} parameters than $\Lambda$CDM+r!) However, entanglement corrections are constrained to be unobservably small, with a 95\%-confidence lower bound on the SUSY breaking scale of $b > 6.46 \times 10^7\ {\rm GeV}$. With respect to tests of predictions:
\begin{itemize}
\item{There is  no suppression of power on large angular scales: power is in fact slightly enhanced on large scales.}
\item{Modulation of the power spectrum is nearly scale invariant, and no observable running of the spectral index is present. Running of the spectral index is of order $\alpha = -0.0006$, consistent with an unmodified Starobinsky $R^2$ potential.}
\item{There are no features induced in the power spectrum, so entanglement corrections do not provide the modulation which would be required to explain the WMAP/Planck cold spot. }
\item{We find no evidence for suppression of $\sigma_8$ relative to the best-fit $\Lambda$CDM+r cosmology.  The best-fit $\Lambda$CDM+r cosmology used in this paper has $\sigma_8 = 0.8299$, and the best-fit Starobinsky model, including entanglement corrections, has $\sigma_8 = 0.8314$, consistent with the Planck $\Lambda$CDM favored value $\sigma_8 = 0.830 \pm 0.015$ \cite{Ade:2015xua}. The $2\sigma$ ``extremal'' Starobinsky model has $\sigma_8 = 0.8439$, which is actually \emph{enhanced} rather than suppressed relative to $\Lambda$CDM. }
\end{itemize}}
\end{itemize}
Appendix \ref{sec:Appendix} contains a table of all parameter fits. 

It is worthwhile to discuss in general the ``concrete predictions'' originally claimed by the authors of Refs. \cite{Holman:2006an,Holman:2006ny}, since several key claims do not survive even cursory scrutiny. For example, the discontinuity in the effective potential claimed to be correlated with voids and the CMB cold spot does not appear to in fact exist: for all physically relevant values of the parameters $V_0$, $\lambda$, and $b$, the modulation $F\left(\phi\right)$ is a smooth function, with no characteristic discontinuities which would explain features in the power spectrum.\footnote{The original numerical code used in Ref. \cite{Holman:2006ny} contained a coding error that resulted in miscalculation of the modulation by nearly two orders of magnitude.} Perhaps more importantly, the form of the effective potential resulting from landscape entanglement is completely dependent on the choice of inflationary potential $V\left(\phi\right)$, which is itself an arbitrary free function. One could just as consistently choose the underlying inflationary potential in the absence of landscape corrections to be the same as the effective potential (\ref{eq:Veff})! In this sense, the landscape model is no more (or less) predictive than single-field inflation itself, and most of the {\it claimed} predictions of the entanglement model turn out not to have been predictions at all. However, any considerations of theoretical consistency are a moot point: even if one takes the claimed predictions at face value, almost all of them are ruled out by Planck. Experiment always supersedes theory, and the model does not match the data.

Finally, we note that the general conclusion about constraints on modulations of the power spectrum, while derived here in the context of a specific model, is a largely \emph{model-independent} statement. Modulations of the primordial power spectrum from entanglement are ruled out precisely because \emph{any} observably large modulations of the primordial power spectrum will introduce substantial tension with the Planck data, relative to a pure power law \cite{Chen:2016vvw,Chen:2016zuu}. While it is undoubtedly possible to {\it a posteriori} engineer a modulated power spectrum which improves the fit to Planck (see, {\it e.g.} the interesting case presented in Ref. \cite{Hazra:2016fkm}), such models will necessarily require multiple tuned parameters and it is therefore questionable in such a case if even a substantially improved fit can be considered favored from a Bayesian perspective. 

\section*{Acknowledgments}

WHK is supported by the National Science Foundation under grant NSF-PHY-1417317. This work was performed in part at the University at Buffalo Center for Computational Research. WHK thanks the University of Valencia, where part of this work was completed, for hospitality and support. WHK thanks Laura Mersini-Houghton and Richard Holman for extensive consultation and collaboration on an earlier version of this work, and Tomo Takahashi for access to original codes. WHK also thanks Sabine Hossenfelder and the referee for discussions which greatly clarified the context of the analysis. 

\bibliographystyle{JHEP}
\bibliography{paper}

\vfill
\appendix
\section{Parameter Table}
\label{sec:Appendix}

This section contains a table of 1-D parameter fits for the models considered in this paper. Error bars are one standard deviation, and upper/lower bounds are 95\% confidence. Parameters listed in parentheses ``()'' are derived parameters calculated from the best-fit values for $V_0$, $b$, and $\lambda$. Also included is the derived best-fit running for the spectral index, $\alpha \equiv d n / d \ln{k}$. Observables are quoted at a pivot scale $k_* = 0.05\ \mathrm{h\ Mpc}^{-1}$.

\bigskip
\renewcommand{\arraystretch}{1.5}
\begin{tabular}{| c | c | c | c |}
\hline
  & $\Lambda$CDM+r & Exponential & Starobinsky \\ \hline \hline
$\Omega_{\rm b} h^2$ & $0.02222473 \pm 0.00016$ & $0.02250676 \pm 0.00015$ & $0.02225708 \pm 0.00014$  \\ \hline
$\Omega_{\rm c} h^2$ &  $0.1199321 \pm 0.0015$ & $0.1155528 \pm 0.0012$ & $ 0.1195799 \pm 0.00095$  \\ \hline
$\theta$ &  $1.040745 \pm 0.00032$ & $1.041232 \pm 0.00032$ & $1.040792 \pm 0.00029$ \\ \hline
$\tau$ & $0.08100027 \pm 0.016$ & $0.1031400 \pm 0.017$ & $0.08644387 \pm 0.015$ \\ \hline
$H_0$ & $67.20348 \pm  0.66$ & $69.1682 \pm 0.55$ & $67.36688 \pm 0.43$ \\ \hline
$\ln{\left(10^{10} A_s\right)}$ & $3.097646 \pm 0.031$ & $(3.14846)$ & $(3.10186)$ \\ \hline
$r$ & $<0.06693180$ (95\%) & $(0.130768)$ & $(0.0032943)$ \\ \hline
$n_{\rm S}$ & $0.9644768 \pm 0.0048$ & $(0.984077)$ & $(0.965991)$ \\ \hline
$\alpha$ & - & $(3.6 \times 10^{-6})$ & $(-0.000061)$ \\ \hline
$\sigma_8$ & (0.8299) & (0.8360) & (0.8314) \\ \hline
$10^{9} V_0/M_{\rm P}^4$ & - & $9.377153 \pm 1.8$ & $ 0.2229236 \pm 0.0068$ \\ \hline
$\log_{10}{b\ ({\rm GeV})}$ & - & $>9.048864$ (95\%) & $>7.811044$ (95\%)\\ \hline
$\lambda$ & - & $0.1261851 \pm 0.012$ & - \\ \hline \hline
Likelihood & $6794.3960$ & $6807.6640$ & $6792.9980$ \\ \hline
\end{tabular} 

\end{document}